\documentclass[draft]{agujournal2019}
\usepackage[inline]{trackchanges} 
\usepackage{soul}
\journalname{JGR: Planets}

\begin{document}

%
%


\title{Low rock mass fraction within trans-Neptunian objects inferred from the spin--orbit evolution of Orcus--Vanth and Salacia--Actaea}

%
%




\authors{S. Arakawa\affil{1}, S. Kamata\affil{2}, H. Genda\affil{3}}

\affiliation{1}{Center for Mathematical Science and Advanced Technology, \\
Japan Agency for Marine-Earth Science and Technology, \\
3173-25, Showa-machi, Kanazawa-ku, Yokohama 236-0001, Japan\\
~}
\affiliation{2}{Department of Earth and Planetary Sciences, Hokkaido University, \\
Kita-10 Nishi-8, Kita-ku, Sapporo 060-0810, Japan\\
~}
\affiliation{3}{Earth-Life Science Institute, Institute of Science Tokyo, \\
2-12-1, Ookayama, Meguro-ku, Tokyo 152-8550, Japan
}





\correspondingauthor{Sota Arakawa}{arakawas@jamstec.go.jp}

\begin{abstract}

Satellites play a crucial role in understanding the formation and evolution of trans-Neptunian objects (TNOs).
The spin--orbit evolution of satellite systems depends on their thermal histories, allowing us to constrain the rock mass fraction within TNOs based on their current spin--orbit states.
In this study, we perform coupled thermal--orbital evolution calculations for two satellite systems around undifferentiated TNOs: Orcus--Vanth and Salacia--Actaea.
Our results demonstrate that the current spin--orbit states of these systems are consistent with a rock mass fraction of approximately 20--30\%.
Additionally, we estimate the organic mass fraction within the TNOs and find that it is comparable to the rock mass fraction.
These findings suggest that the chemical composition of TNOs closely resembles that of comets.

\end{abstract}





%
%

\section{Introduction}

Satellites are of great importance in understanding the formation and evolution of trans-Neptunian objects (TNOs).
For instance, the semimajor axis and orbital period of a satellite system enable us to determine the system's mass \cite<e.g.,>[]{2008ssbn.book..345N}.
The obliquity and eccentricity provide clues about the satellite's origin and its subsequent tidal evolution history \cite<e.g.,>[]{2021AJ....162..226A, 2021PSJ.....2....4R}.
Deviations of the mutual orbit from Keplerian motion reveal information about the objects' shapes \cite<e.g.,>[]{2024AJ....167..144P}.
Additionally, by combining the system's mass with the radius of the orbiting bodies, we can estimate their bulk density, which offers key insights into their internal structure \cite<e.g.,>[]{2017AJ....154...19B, 2024A&A...684A..50K}.
It is known that the majority of TNOs with radii comparable to or larger than $300~{\rm km}$ host one or more satellites orbiting the primary \cite<e.g.,>[]{2017ApJ...838L...1K}.
Furthermore, many $100~{\rm km}$-sized TNOs exist as binary (or triplet) systems with nearly equal primary-to-secondary mass ratios \cite<e.g.,>[]{2020tnss.book..201N}.

Several mechanisms have been proposed to explain the origin of satellite and binary systems, including giant impacts \cite<e.g.,>[]{2019NatAs...3..802A, 2021psnh.book..475C, 2025NatGe..18...37D}, rotational fission/ejection \cite<e.g.,>[]{2012MNRAS.419.2315O, 2022PSJ.....3..225N}, dynamical capture \cite<e.g.,>[]{2008ApJ...673.1218S, 2011PASJ...63.1331K}, and direct formation during the accretion phase \cite<e.g.,>[]{2019NatAs...3..808N, 2020A&A...643A..55R}.
Following the formation of a satellite system, its spin--orbit state evolves due to tidal interactions, which depend on the thermal history of the objects \cite<e.g.,>[]{2021AJ....162..226A, 2022Icar..37614871B}.
The main heat sources of TNOs are radioactive nuclei such as $^{40}$K and $^{238}$U \cite<e.g.,>[]{2011Icar..216..426R}; thus, the temporal evolution of their internal temperature is determined by the initial abundances of these nuclei, which are proportional to the rock mass fraction within the object.
In other words, the current spin--orbit states can provide constraints on the rock mass fraction within TNOs.

The internal structure of TNOs is another critical factor influencing their thermal and orbital evolution.
To understand tidal evolution, it is essential to know the strength of tidal dissipation within these objects.
However, for ice--rock differentiated objects, multiple factors contribute to significant uncertainties.
For example, there is no consensus on the thickness of subsurface oceans, which depends on various assumptions, such as the presence/absence of thermally insulating layers on the ocean \cite<e.g.,>[]{2019NatGe..12..407K, 2020P&SS..18104828K, 2024NatAs...8..748B}.
The temperature structure of the ice shell is also affected by the presence of insulating layers \cite<e.g.,>[]{2019NatGe..12..407K}, which drastically alters the strength of tidal dissipation.
Moreover, whether the core of a differentiated TNO is dissipative remains debated.
If the core consists of a mixture of solid rocks and liquid water, as hypothesized for the icy satellite Enceladus, the core would be dissipative \cite<e.g.,>[]{2015Icar..258...54R, 2022JGRE..12707117R}.
Although dissipation strength can be calculated using a poroviscoelastic gravitational model \cite<e.g.,>[]{2023JGRE..12807700K}, the model requires parameters such as permeability, which remain poorly understood.
In comparison, undifferentiated bodies present a simpler scenario, making spin--orbit states of satellite systems around undifferentiated TNOs particularly useful for constraining the rock mass fraction within TNOs.

\begin{figure}
\centering
\includegraphics[width = \columnwidth]{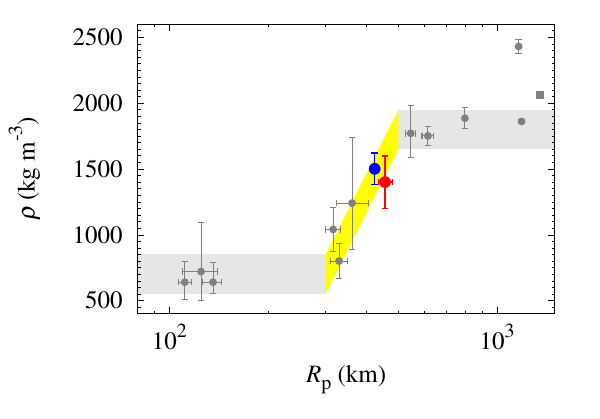}
\caption{
The size--density relationship of TNOs hosting satellite(s), with system masses determined by their mutual orbits.
The red circle represents the radius and density of Orcus \cite{2018AJ....156..164B}, while the blue circle denotes those of Salacia \cite{2019Icar..334...62G}.
Gray circles indicate the values for other TNOs with $R_{\rm p} > 100~{\rm km}$: Pluto \cite{2015Sci...350.1815S}, Eris \cite{2021Icar..35514130H}, Haumea \cite{2017Natur.550..219O}, Gonggong \cite{2019Icar..334....3K}, Quaoar \cite{2024A&A...684A..50K}, Varda \cite{2015Icar..257..130G}, 2002 UX$_{25}$ \cite{2017AJ....154...19B}, G!k\'{u}n$\|$'h\`{o}md\'{i}m\`{a} \cite{2019Icar..334...30G}, Lempo \cite{2012A&A...541A..93M}, Sila \cite{2012Icar..220...74G}, and Ceto \cite{2012A&A...541A..92S}.
The gray square represents data for Triton as a reference \cite{2000Icar..148..587T}.
The gray and yellow shaded regions indicate the density ranges.
We note that Makemake and Sedna are not plotted here because their masses are difficult to constrain.
Sedna has no known satellites, and the orbit of Makemake's satellite remains highly uncertain \cite{2016ApJ...825L...9P}.
}
\label{fig:size-density}
\end{figure}

In this study, we focus on the satellite systems of two TNOs: (90482) Orcus and (120347) Salacia.
These TNOs each host a known satellite, Vanth and Actaea, respectively, and their radii and system masses have been constrained through recent precise observations \cite<e.g.,>[]{2018AJ....156..164B, 2019Icar..334...62G, 2019Icar..319..657S}.
From their radii and masses, we find that these TNOs have low but nonzero porosity, suggesting an undifferentiated nature (see Figure \ref{fig:size-density}).
We perform numerical simulations of coupled thermal--orbital evolution for these systems under various rock mass fraction settings (Section \ref{sec:results}).
This is the first study to constrain the composition of TNOs based not only on their densities but on their spin--orbit evolution.
Our findings indicate that the current spin--orbit states of these two systems can be reproduced when the rock mass fraction is approximately 20--30\%. 
We also provide a preliminary estimate of the organic mass fraction within TNOs and find that the organic mass fraction may be comparable to or even exceed the rock mass fraction (Section \ref{sec:organic}).
The estimated composition of TNOs aligns with that of comets, highlighting the potentially significant role of organic materials in the formation and evolution of planetary objects in the outer solar system.

\section{Size--density relationship of TNOs}
\label{sec:size-density}

For a satellite system whose primary radius, $R_{\rm p}$, is determined from radiometry and/or occultation measurements, the bulk density of the primary, $\rho$, can be estimated.
Recent precise measurements of the radii of TNOs sized $100$--$1000~{\rm km}$ have revealed a clear size--density relationship \cite<e.g.,>[]{2024A&A...684A..50K}.
Figure \ref{fig:size-density} illustrates the size--density relationship for TNOs that host satellite(s) and whose system masses are determined by their mutual orbits.
For large TNOs with $R_{\rm p} \ge 500~{\rm km}$, the density lies within the range of ${\left( 1800 \pm 150 \right)}~{\rm kg}~{\rm m}^{-3}$ and shows little dependence on $R_{\rm p}$.
This suggests that these objects are consolidated without porosity.
In contrast, small TNOs with $R_{\rm p} \le 300~{\rm km}$ typically exhibit a density of $\rho \approx {\left( 700 \pm 150 \right)}~{\rm kg}~{\rm m}^{-3}$.
Assuming these small objects share the same composition as large TNOs, their low densities imply significant porosity.
The porosity, $\varepsilon$, is defined as
\begin{equation}
\rho = {\left( 1 - \varepsilon \right)} \rho_{\rm mat},
\end{equation}
where $\rho_{\rm mat} \approx 1800~{\rm kg}~{\rm m}^{-3}$ represents the average material density.
For $\rho \approx 700~{\rm kg}~{\rm m}^{-3}$, the porosity is approximately $\varepsilon \approx 60\%$ .

For mid-sized TNOs with $300~{\rm km} < R_{\rm p} < 500~{\rm km}$ (yellow shaded region), a clear size--density relationship is observed.
In this regime, $\rho$ increases with $R_{\rm p}$ likely due to reduced porosity \cite<e.g.,>[]{2009JGRE..114.9004Y, 2019Icar..326...10B}.
Thus, these mid-sized TNOs would have small but nonzero porosity.
Fully differentiated bodies are expected to be nonporous, as their entire interiors have undergone melting events.
Thus, the finite porosity of mid-sized TNOs provides evidence that a large fraction of their interiors would remain undifferentiated.

For both Orcus and Salacia, recent observations reported a density of $\rho \approx 1500~{\rm kg}~{\rm m}^{-3}$ \cite{2018AJ....156..164B, 2019Icar..334...62G}.
Assuming $\rho_{\rm mat} \approx 1800~{\rm kg}~{\rm m}^{-3}$, the porosity is estimated as $\varepsilon \approx 17\%$.
Given the small estimated $\varepsilon$, it can be assumes that the material properties (e.g., viscosity, shear modulus, and thermal conductivity) of the interiors of Orcus and Salacia would be on the same order of magnitude as those of consolidated materials.
For TNOs smaller than Salacia, in contrast, it is essential to take the effects of porosity into consideration when evaluating the material properties of their interior.

We note that the second largest TNO, Eris, has an exceptionally high density of $\rho \approx 2430~{\rm kg}~{\rm m}^{-3}$ \cite{2021Icar..35514130H}.
Its satellite, Dysnomia, has a radius of $\approx 310~{\rm km}$ and a density of $< 1200~{\rm kg}~{\rm m}^{-3}$ \cite{2023PSJ.....4..193B}.
One possible origin for the Eris--Dysnomia system is a giant impact between differentiated progenitors, resulting in the merging of dense rocky cores and the ejection of icy fragments as moonlets \cite<see Figure S1 of>[]{2019NatAs...3..802A}.
The low density of Dysnomia is consistent with the scenario in which it formed as an icy fragment during the giant impact event, although we cannot rule out the possibility that Dysnomia is a porous satellite.
The dual-synchronous state of the system \cite{2023A&A...669L...3S, 2023PSJ.....4..115B} implies that Eris's interior was tidally dissipative \cite{2023SciA....9I9201N}.
The high rock mass fraction of Eris inferred from its density is consistent with this tidal evolution scenario.

\section{Models}
\label{sec:model}

In Section \ref{sec:model}, we introduce the numerical model.
We calculate the coupled thermal--orbital evolution of two trans-Neptunian satellite systems, Orcus--Vanth and Salacia--Actaea, using a viscoelastic rheological framework.
Table \ref{table:system} summarizes the fundamental parameters of the current Orcus--Vanth and Salacia--Actaea systems.
For simplicity, we assume that the densities of the primaries and secondaries are equal.

\begin{table}
\caption{Fundamental parameters of the current satellite systems: the radii of the primary and secondary ($R_{\rm p}$ and $R_{\rm s}$), the system mass ($M_{\rm sys}$), the current semimajor axis ($a_{\rm obs}$), and the current spin period of the primary ($P_{\rm p, obs}$).  
Data are taken from \citeA{2018AJ....156..164B}, \citeA{2019Icar..334...62G}, \citeA{2019Icar..319..657S}, and \citeA{2020EPSC...14..516K} for the Orcus--Vanth system, and from \citeA{2019Icar..334...62G} and \citeA{2014A&A...569A...3T} for the Salacia--Actaea system.
We note that \citeA{2020EPSC...14..516K} is a conference abstract and that the spin period of Orcus is preliminary.
}
\centering
\begin{tabular}{l c c c c c}
\hline
System              & $R_{\rm p}$ (km)   & $R_{\rm s}$ (km)   & $M_{\rm sys}$ (kg)   & $a_{\rm obs}$ (km)   & $P_{\rm p, obs}$ (h) \\
\hline
Orcus--Vanth        & $460$              & $220$              & $6.3 \times 10^{20}$ & $9.0 \times 10^{3}$  & $7$                  \\
Salacia--Actaea     & $420$              & $140$              & $4.9 \times 10^{20}$ & $5.7 \times 10^{3}$  & $6.5$                \\
\hline
\end{tabular}
\label{table:system}
\end{table}

\subsection{Rheology of ice--refractory mixture}

The tidal evolution of satellite systems significantly depends on the viscoelastic response of the bodies, making the temperature-dependent viscosity of the ice--refractory mixture a key parameter for understanding the spin--orbit evolution of satellite systems of mid-sized TNOs.
The interiors of these objects are undifferentiated and consist of a mixture of micron-sized refractory (rock and organic materials) grains and matrix ice.
The grain size of ice crystals would be maintained at the micron scale due to the Zener pinning effect \cite<e.g.,>[]{2009JMPeS.104..301K}.
Additionally, convective flow itself can inhibit ice crystal grain growth \cite<e.g.,>[]{2022PCM....49...28C}.

The effective viscosity of the mixture, $\eta$, increases with the volume fraction of refractory grains, $\phi_{\rm grain}$  \cite{1906AnP...324..289E}.
It can be approximated as \cite<e.g.,>[]{1959JRheo...3..137K}:
\begin{equation}
\eta \approx {\left( 1 - \frac{\phi_{\rm grain}}{\phi_{\rm max}} \right)}^{- 2.5 \phi_{\rm max}} \eta_{\rm ice},
\label{eq:eta_mix}
\end{equation}
where $\eta_{\rm ice}$ is the viscosity of pure ice, and $\phi_{\rm max}$ is the volume fraction of refractory grains at the jamming point.
For monodispersed spheres, $\phi_{\rm max} \approx 0.64$ \cite<e.g.,>[]{1983PhRvA..27.1053B}.
However, \citeA{2020JGRE..12506519G} reported that $\phi_{\rm max} = 0.8$--$0.9$ if the grain size follows a power-law distribution similar to the boulders on asteroid Ryugu. 
Equation (\ref{eq:eta_mix}) suggests that $\eta / \eta_{\rm ice} = 6$--$12$ for $\phi_{\rm grain} = 0.5$, indicating a one-order-of-magnitude increase in viscosity compared to pure ice.

The creep behavior of ice have been extensively studied through laboratory experiments \cite<e.g.,>[]{2001JGR...10611017G, 2006Sci...311.1267K, 2010SSRv..153..273D, 2020Icar..33513401N}.
Diffusion creep is thought to control the viscosity of icy bodies smaller than 1000 km in diameter \cite<e.g.,>[]{2005JGRE..11012005B, 2006Icar..183..435M}.
For polycrystalline ice, the viscosity is given by \cite{2001JGR...10611017G}:
\begin{equation}
\eta_{\rm ice} = A {d_{\rm ice}}^{2} \exp{\left[ \frac{E_{\rm a}}{R_{\rm gas} T_{\rm ref}} {\left( \frac{T_{\rm ref}}{T} - 1 \right)} \right]},
\label{eq:eta_ice}
\end{equation}
where $A = 9 \times 10^{20}~{\rm Pa}~{\rm s}~{\rm m}^{-2}$ is a material constant, $d_{\rm ice}$ is the grain size of ice crystals, $E_{\rm a} = 60~{\rm kJ}~{\rm mol}^{-1}$ is the activation energy for volume diffusion, $R_{\rm gas}$ is the gas constant, and $T_{\rm ref} = 273~{\rm K}$ is the reference temperature.
for $d_{\rm ice} \approx 1~{\mu}{\rm m}$, the ice viscosity is $\eta_{\rm ice} \approx 10^{9}~{\rm Pa}~{\rm s}$, while for $d_{\rm ice} \approx 0.1~{\mu}{\rm m}$, $\eta_{\rm ice} \approx 10^{7}~{\rm Pa}~{\rm s}$.
By combining Equations (\ref{eq:eta_mix}) and (\ref{eq:eta_ice}), the temperature dependence of $\eta$ is expressed as:
\begin{equation}
\eta = \eta_{\rm ref} \exp{\left[ \frac{E_{\rm a}}{R_{\rm gas} T_{\rm ref}} {\left( \frac{T_{\rm ref}}{T} - 1 \right)} \right]},
\label{eq:eta_ref}
\end{equation}
where $\eta_{\rm ref}$ is treated as a parameter.

The complex shear modulus, $\tilde{\mu}$, is a key parameter for describing the deformation of ice--refractory mixtures under periodic forcing.
The Andrade model \cite{1910RSPSA..84....1A} is widely used to characterize the anelastic and viscoelastic response of ice \cite<e.g.,>[]{1955RSPSA.228..519G, 2013Icar..226...10S, 2019NatAs...3..543N, 2020Icar..34313610G}, and we apply the model to ice--refractory mixtures.
The complex shear modulus is expressed as:
\begin{equation}
\tilde{\mu} = \frac{1}{\tilde{J}},
\end{equation}
where $\tilde{J}$ is the complex creep function, defined as follows \cite<e.g.,>[]{2012ApJ...746..150E}:
\begin{equation}
\tilde{J} = {\left[ \frac{1}{\mu} + \frac{\cos{\left( \alpha_{\rm A} \pi / 2 \right)}~\Gamma{\left( \alpha_{\rm A} + 1 \right)}}{\mu {\left( \tau_{\rm A} \omega \right)}^{\alpha_{\rm A}}} \right]} + i {\left[ \frac{1}{\eta \omega} + \frac{\sin{\left( \alpha_{\rm A} \pi / 2 \right)}~\Gamma{\left( \alpha_{\rm A} + 1 \right)}}{\mu {\left( \tau_{\rm A} \omega \right)}^{\alpha_{\rm A}}} \right]},
\end{equation}
where $\mu$ is the the static shear modulus, $\omega$ is the forcing frequency, $\tau_{\rm A} = \eta / \mu$ is the relaxation timescale, and $\alpha_{\rm A} = 0.33$ is the Andrade exponent \cite<e.g.,>[]{2013Icar..226...10S, 2020Icar..34313610G}.

In our fiducial model (Model {\bf F}), we set $\eta_{\rm ref} = 10^{10}~{\rm Pa}~{\rm s}$ and $\mu = 3.3~{\rm GPa}$ (see Table \ref{table:model}).
We note that there is uncertainty in both $\eta_{\rm ref}$ and $\mu$.
\citeA{2009ApJ...691...54G} pointed out that $\mu$ for bodies with pores is lower than that for nonporous bodies \cite<see also>[]{1990RSPSA.430..105G, 2019Icar..321..715N}.
The value of $\eta_{\rm ref}$ depends on the volume fraction of refractory grains and the grain size of ice crystals.
Thus, we also investigate cases with a low static shear modulus ($\mu = 0.33~{\rm GPa}$; Model {\bf M}) and with a low reference viscosity ($\eta_{\rm ref}= 10^{8}~{\rm Pa}~{\rm s}$; Model {\bf H}).

\begin{table}
\caption{Parameter values adopted in four models.}
\centering
\begin{tabular}{l c c c}
\hline
Model                             & $\mu$ (GPa)   & $k_{\rm th}$ (${\rm W}~{\rm m}^{-1}~{\rm K}^{-1}$)   & $\eta_{\rm ref}$ (Pa s)   \\
\hline
{\bf F} (fiducial)                & $3.3$         & $3$                                                  & $10^{10}$                 \\
{\bf M} (low $\mu$)               & $0.33$        & $3$                                                  & $10^{10}$                 \\
{\bf K} (low $k_{\rm th}$)        & $3.3$         & $1$                                                  & $10^{10}$                 \\
{\bf H} (low $\eta_{\rm ref}$)    & $3.3$         & $3$                                                  & $10^{8}$                  \\
\hline
\end{tabular}
\label{table:model}
\end{table}

\subsection{Internal temperature structure}
\label{sec:internal}

We perform calculations of coupled thermal--orbital evolution using a simplified thermal structure model, as described below.
The internal temperature structure within the primary body is expressed as follows:
\begin{equation}
T {( r )} = \left\{ 
\begin{array}{ll}
T_{\rm c},   &   ( 0 \le r < R_{\rm p} - \Delta )  \\
T_{\rm s} + {\left( T_{\rm c} - T_{\rm s} \right)} {\left( R_{\rm p} - r \right)} / \Delta, & ( R_{\rm p} - \Delta \le r \le R_{\rm p} )
\end{array}
\right.
\label{eq:T_r}
\end{equation}
where $T_{\rm c}$ and $T_{\rm s}$ are the temperatures at the center and the surface, respectively, $r$ is the distance from the center, and $\Delta$ denotes the thickness of the conductive lid.
We set $T_{\rm s} = 40~{\rm K}$, consistent with assumptions made in previous studies on the thermal evolution of TNOs \cite<e.g.,>[]{2019NatGe..12..407K, 2021AJ....162..226A}.

When $T_{\rm c}$ exceeds the critical point, stagnant lid convection is expected to occur on TNOs \cite<e.g.,>[]{2014PEPI..229...40D, 2018GeoFr...9..103S}.
We assume that $\Delta$ is given by
\begin{equation}
\Delta = R_{\rm p} / {\rm Nu},
\end{equation}
where ${\rm Nu}$ is the Nusselt number, which depends on the Frank--Kamenetskii parameter, $\Theta$, and the Rayleigh number, ${\rm Ra}$ \cite{2005PEPI..149..361R}.
The Nusselt number is expressed as
\begin{equation}
{\rm Nu} = \max{\left[ 0.67 \Theta^{- 4/3} {\rm Ra}^{1/3}, 1 \right]},
\label{eq:Nu}
\end{equation}
where ${\rm Nu} > 1$ indicates that convection is active within the body.
The parameters $\Theta$ and ${\rm Ra}$, both functions of $T_{\rm c}$, are given as follows:
\begin{eqnarray}
\Theta   & = & \frac{E_{\rm a} {\left( T_{\rm c} - T_{\rm s} \right)}}{R_{\rm gas} {T_{\rm c}}^{2}}, \\
{\rm Ra} & = & \frac{\alpha_{\rm e} \rho g {\left( T_{\rm c} - T_{\rm s} \right)}}{\kappa \eta_{\rm c}} {R_{\rm p}}^{3},
\end{eqnarray}
where $\alpha_{\rm e} = 1 \times 10^{-4}~{\rm K}^{-1}$ is the thermal expansion coefficient, $g = \mathcal{G} M_{\rm p} / {R_{\rm p}}^{2}$ is the surface gravity, $\mathcal{G}$ is the gravitational constant, and $\eta_{\rm c}$ is the viscosity at $T_{\rm c}$ (see Equation (\ref{eq:eta_ref})).
The thermal diffusivity, $\kappa$, is defined as $\kappa = k_{\rm th} / {( \rho c )}$, where $k_{\rm th}$ is the thermal conductivity and $c = 1 \times 10^{3}~{\rm J}~{\rm kg}^{-1}~{\rm K}^{-1}$ is the specific heat capacity.

In our fiducial model, we set $k_{\rm th} = 3~{\rm W}~{\rm m}^{-1}~{\rm K}^{-1}$ and ignore its temperature dependence for simplicity (see Table \ref{table:model}).
We note that $k_{\rm th}$ depends on the composition and decreases with increasing volume fraction of organics \cite<e.g.,>[]{2020MNRAS.497.1166A, 2023E&PSL.61218172R}.
Furthermore, $k_{\rm th}$ for porous materials is lower than that for nonporous materials due to reduced heat path \cite<e.g.,>[]{1983JGR....88.9513Y, 2019Icar..324....8A, 2022AsBio..22.1047N}.
Thus, we also investigate the case with a low thermal conductivity ($k_{\rm th} = 1~{\rm W}~{\rm m}^{-1}~{\rm K}^{-1}$; Model {\bf K}).


\subsection{Thermal evolution}
\label{sec:thermal}

We employ the simplifying assumption for thermal evolution as assumed in \citeA{2021AJ....162..226A}.
The temporal evolution of $T_{\rm c}$ is calculated by considering the energy budget of the entire body:
\begin{equation}
\frac{{\rm d}T_{\rm c}}{{\rm d}t} = \frac{1}{M_{\rm p} c} {\left( \dot{Q} - 4 \pi {R_{\rm p}}^{2} F \right)},
\end{equation}
where $\dot{Q}$ is the heat generation rate within the body, and $F$ is the heat flux at the surface.
The heat flux $F$ is expressed as:
\begin{equation}
F = k_{\rm th} \frac{T_{\rm c} - T_{\rm s}}{\Delta}.
\end{equation}
The primary source of heat generation is the decay heat from radioactive isotopes.
The heat generation rate $\dot{Q}$ is given by:
\begin{equation}
\dot{Q} = f_{\rm rock} M_{\rm p} \sum_{i} H_{i} \exp{\left( - \frac{t}{\tau_{i} / \ln{2}} \right)},
\label{eq:Q}
\end{equation}
where $f_{\rm rock}$ is the rock mass fraction, $H_{i}$ is the decay heat of radioactive nuclide $i$ per unit mass of rock at $t = 0$, and $\tau_{i}$ is the half-life time of species $i$.
We consider four long-lived radioactive species as heat sources: ${}^{40}{\rm K}$, ${}^{232}{\rm Th}$, ${}^{235}{\rm U}$, and ${}^{238}{\rm U}$ \cite{2011Icar..216..426R}.
The decay data ($\tau_{i}$ and $H_{i}$) are listed in Table \ref{table:heat}.
We assume that satellite systems formed within the first 100 million years of the solar system \cite<e.g.,>[]{2019NatAs...3..802A} and that the initial abundances of long-lived radioactive species were nearly identical to those at the formation time of calcium--aluminum-rich inclusions.
We do not consider the decay heat of short-lived $^{26}{\rm Al}$.
The contribution of $^{26}{\rm Al}$ is negligible if satellite systems formed more than $10~{\rm Myr}$ after the formation of calcium--aluminum-rich inclusions \cite<e.g.,>{2015A&A...584A.117N}.
In this study, $f_{\rm rock}$ is treated as a parameter to investigate its impact on the spin--orbit evolution of satellite systems.
Since $\dot{Q}$ is proportional to $f_{\rm rock}$, calculations with larger $f_{\rm rock}$ result in higher $T_{\rm c}$.

\begin{table}
\caption{Radioactive species and decay data \cite{2011Icar..216..426R}.
We assume a chemical composition corresponding to CI carbonaceous chondrites.
We use $H_{i}$ values corresponding to the formation time of calcium--aluminum-rich inclusions.}
\centering
\begin{tabular}{c c c}
\hline
Element $i$         & $\tau_{i}$ (Gyr)   & $H_{i}$ ($10^{-12}~{\rm W}~{\rm kg}^{-1}$)    \\
\hline
$^{40}$K            & $1.28$             & $21.5$               \\
$^{232}$Th          & $14.0$             & $1.02$               \\
$^{235}$U           & $0.703$            & $3.07$               \\
$^{238}$U           & $4.47$             & $1.88$               \\
\hline
\end{tabular}
\label{table:heat}
\end{table}

\subsection{Spin--orbit evolution}

We calculate the secular spin--orbit evolution after the formation of trans-Neptunian satellite systems.
We assume that the eccentricity of the mutual orbit remains negligibly small during tidal evolution.
Although their initial mutual orbits may have non-zero eccentricity if they were formed via giant impacts \cite<e.g.,>[]{2019NatAs...3..802A}, the timescale for eccentricity damping is orders of magnitude shorter than those for orbit expansion and tidal despinning \cite<e.g.,>[]{2021AJ....162..226A}.

The temporal evolution of the semimajor axis, $a$, is given by the following equation \cite{1999ssd..book.....M}:
\begin{equation}
\frac{{\rm d}a}{{\rm d}t} = 3 \sqrt{\mathcal{G} M_{\rm sys}} \frac{M_{\rm s}}{M_{\rm p}} {R_{\rm p}}^{5} a^{- 11/2}~{{\rm Im} {\left( k_{2} \right)}},
\label{eq:dadt}
\end{equation}
where ${\rm Im} {\left( k_{2} \right)}$ is the imaginary part of the complex Love number, $k_{2}$ (see Section \ref{sec:k2}).

The spin period of the primary, $P_{\rm p}$, is given by $P_{\rm p} = 2 \pi / \dot{\theta}_{\rm p}$, where  $\dot{\theta}_{\rm p}$ is the spin angular velocity of the primary.
We calculate $\dot{\theta}_{\rm p}$ using the conservation equation for the total angular momentum of the system, $L$:
\begin{equation}
L = I_{\rm p} \dot{\theta}_{\rm p} + \frac{M_{\rm p} M_{\rm s}}{M_{\rm sys}} n a^{2} = {\rm constant},
\label{eq:L}
\end{equation}
where $I_{\rm p} = {( 2 / 5 )} M_{\rm p} {R_{\rm p}}^{2}$ is the moment of inertia of the primary, and $n = \sqrt{\mathcal{G} M_{\rm sys} / a^{3}}$ is the mean motion.
We note that the spin angular momentum of the secondary is negligible in comparison to $L$.
The tidal forcing frequency, $\omega_{\rm f}$, is given by $\omega_{\rm f} = 2 \dot{\theta}_{\rm p} - 2 n$, and the corresponding forcing period is $P_{\rm f} = 2 \pi / \omega_{\rm f}$.

We integrate Equation (\ref{eq:dadt}) from $t = 0$ to $4.5~{\rm Gyr}$.
Assuming that satellites are formed via giant impacts, \citeA{2021AJ....162..226A} found that the semimajor axis lies within the range of $3 R_{\rm p} \le a \le 8 R_{\rm p}$ after the orbit of satellites are circularized.
In this study, we set $a = 6 R_{\rm p}$ at $t = 0$, and the initial value of $P_{\rm p}$ is calculated from Equation (\ref{eq:L}): $P_{\rm p, ini} = 3.1~{\rm h}$ for Orcus and $5.1~{\rm h}$ for Salacia.
These initial spin periods are longer than the critical value for rotational instability \cite<e.g.,>[]{2000Icar..148...12P, 2021arXiv211015258P}.
We note that the final semimajor axis is nearly independent of the initial semimajor axis.

\subsection{Love number}
\label{sec:k2}

We calculate $k_{2}$ using a viscoelasto-gravitational theory for the periodic deformation of spherical bodies \cite<see Section 2 of>[]{2015JGRE..120.1528K}.
We consider the radial temperature structure within Orcus and Salacia (Equation (\ref{eq:T_r})).
Thus, $\eta$ varies with $r$, which in turn affects $\tilde{\mu}$.
Using the numerical code {\bf LNTools} \cite{2023zndo...7804175K}, we calculate $k_{2}$ as a function of $P_{\rm f}$ (or $\omega_{\rm f}$) and $T_{\rm c}$:
\begin{equation}
k_{2} = k_{2} {\left( P_{\rm f}, T_{\rm c} \right)}.
\end{equation}

\begin{figure}
\centering
\includegraphics[width = \columnwidth]{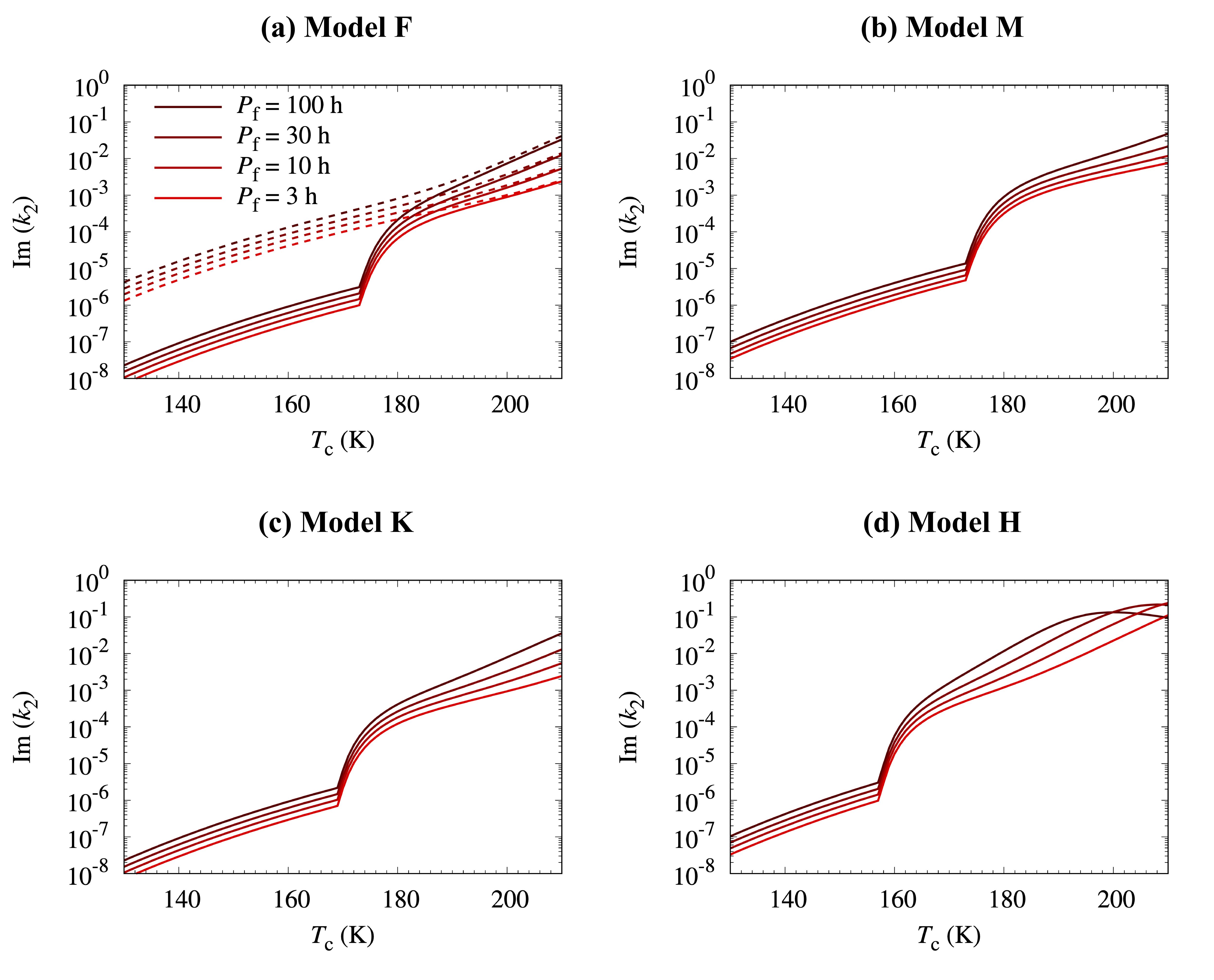}
\caption{
Temperature- and period-dependence of ${\rm Im} {\left( k_{2} \right)}$ for Orcus.
(a) Fiducial model ({\bf F}).
(b) Low $\mu$ model ({\bf M}).
(c) Low $k_{\rm th}$ model ({\bf K}).
(d) Low $\eta_{\rm ref}$ model ({\bf H}).
The ${\rm Im} {\left( k_{2} \right)}$ value sharply increases at the critical $T_{\rm c}$ for convection (see Equation (\ref{eq:Nu})).
The dashed lines in Panel (a) represent the ${\rm Im} {\left( k_{2} \right)}$ values for the homogeneous temperature case (Equation (\ref{eq:k2_homo})).
}
\label{fig:k2}
\end{figure}

Figure \ref{fig:k2} shows the temperature- and period-dependence of ${\rm Im} {\left( k_{2} \right)}$ for Orcus.
For the fiducial model (solid lines in Figure \ref{fig:k2}(a)), ${\rm Im} {\left( k_{2} \right)}$ increases with increasing $T_{\rm c}$ and $P_{\rm f}$.
The value of ${\rm Im} {\left( k_{2} \right)}$ sharply increases at $T_{\rm c} \approx 173~{\rm K}$.
This critical value of $T_{\rm c}$ corresponds to the onset of convection (see Equation (\ref{eq:Nu})).

For calculations of the complex Love number, accounting for the radial temperature structure is crucial.
When $\tilde{\mu}$ is homogeneous within the body, $k_{2}$ is given by the following expression \cite<e.g.,>[]{1973RvGSP..11..767P}:
\begin{equation}
k_{2} = \frac{3}{2} {\left( 1 + \frac{19 \tilde{\mu}}{2 g \rho R_{\rm p}} \right)}^{-1}.
\label{eq:k2_homo}
\end{equation}
The dashed lines in Figure \ref{fig:k2}(a) show the value of ${\rm Im} {\left( k_{2} \right)}$ for the case where the temperature within the body is constant (i.e., $T {( r )} = T_{\rm c} = {\rm constant}$).
We find that ${\rm Im} {\left( k_{2} \right)}$ could be overestimated by orders of magnitude for $T_{\rm c} \le 180~{\rm K}$ if we neglect the radial temperature structure.
In contrast, for $T_{\rm c} \ge 200~{\rm K}$, the assumption of homogeneous internal temperature provides a good approximation for the calculation of $k_{\rm 2}$.

The value of ${\rm Im} {\left( k_{2} \right)}$ and its dependence on $T_{\rm c}$ and $P_{\rm f}$ differ among the models.
For the low $\mu$ model (Model {\bf M}; Figure \ref{fig:k2}(b)), ${\rm Im} {\left( k_{2} \right)}$ is larger than that for the fiducial model.
For the low $k_{\rm th}$ model (Model {\bf K}; Figure \ref{fig:k2}(c)), the critical $T_{\rm c}$ for convection is approximately $169~{\rm K}$, which is slightly lower than that for the fiducial model ($\approx 173~{\rm K}$).
However, ${\rm Im} {\left( k_{2} \right)}$ is nearly identical to that for the fiducial model, except in the temperature range of $169~{\rm K} < T_{\rm c} < 180~{\rm K}$.
Finally, for the low $\eta_{\rm ref}$ model (Model {\bf H}; Figure \ref{fig:k2}(d)), the critical $T_{\rm c}$ for convection ($\approx 157~{\rm K}$) is lower than that for the fiducial model, and ${\rm Im} {\left( k_{2} \right)}$ is larger than that for the fiducial model when $T_{\rm c}$ and $P_{\rm f}$ are fixed.

\section{Results}
\label{sec:results}

We present numerical results of the coupled thermal--orbital evolution for the Orcus--Vanth and Salacia--Actaea systems in Sections \ref{sec:OV} and \ref{sec:SA}, respectively.
We treat the rock mass fraction ($f_{\rm rock}$; Equation (\ref{eq:Q})) and $T_{\rm c}$ at $t = 0$ ($T_{\rm c, ini}$) as parameters.
Additionally, we compare the numerical results with the current spin--orbit state of these systems.

\subsection{Orcus--Vanth}
\label{sec:OV}

\begin{figure}
\centering
\includegraphics[width = \columnwidth]{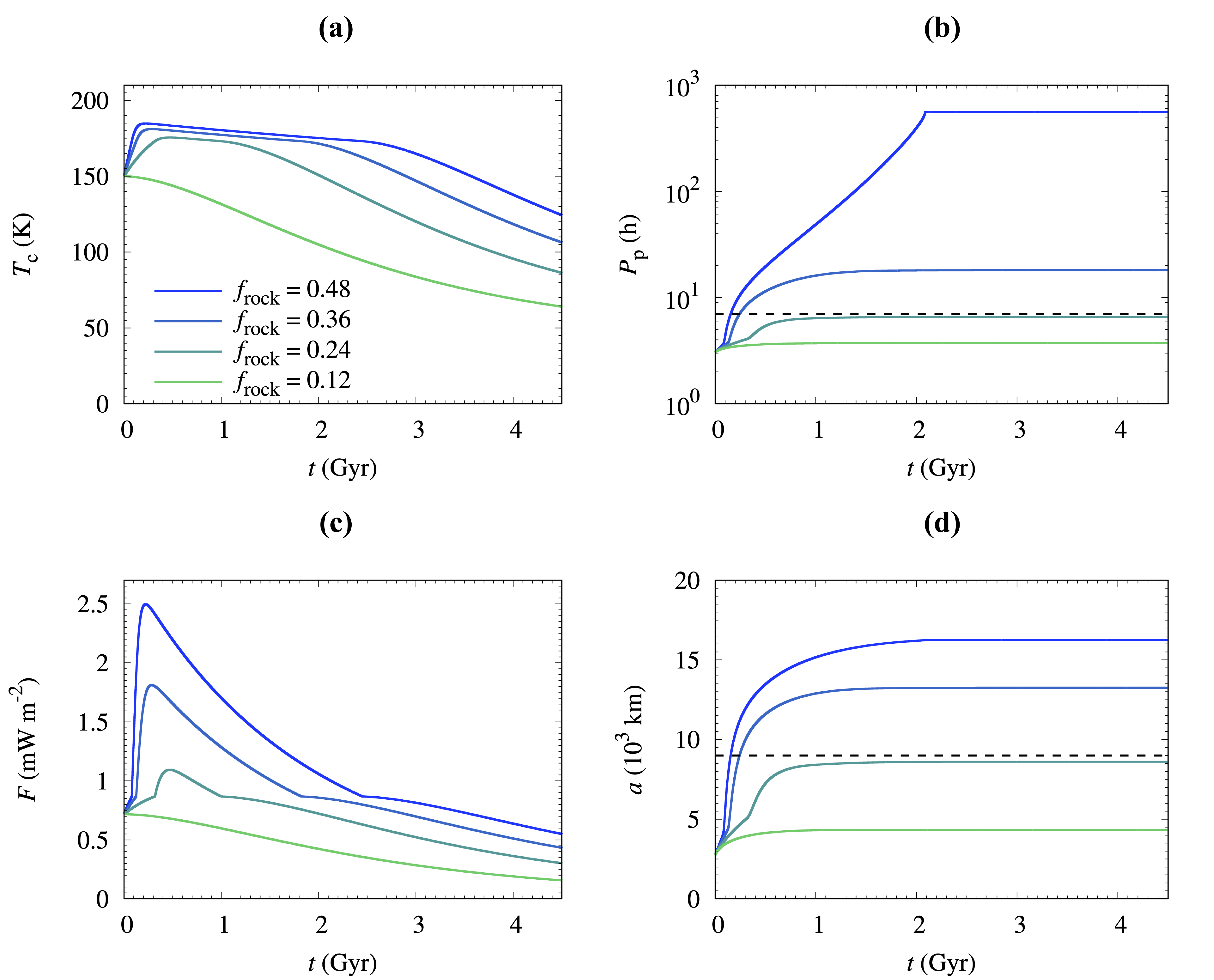}
\caption{
Thermal--orbital evolution of Orcus--Vanth and its dependence on $f_{\rm rock}$.
Here, we set $T_{\rm c, ini} = 150~{\rm K}$ and results for Model {\bf F} are presented.
The dashed lines represent the observed value for the current system (Table \ref{table:system}).
(a) Temperature at the center of Orcus.
(b) Spin period of Orcus.
(c) Heat flux at the surface of Orcus.
(d) Semimajor axis of the mutual orbit.
}
\label{fig:OV_F}
\end{figure}

Figure \ref{fig:OV_F} shows the time evolution of the Orcus--Vanth system for the fiducial model.
Here, we set $T_{\rm c, ini} = 150~{\rm K}$ and examine the dependence on $f_{\rm rock}$.
As shown in Figure \ref{fig:OV_F}(a), $T_{\rm c}$ increases with $f_{\rm rock}$ when the other parameters are fixed.
A plateau in the temperature curve appears for $f_{\rm rock} = 0.24$, $0.36$, and $0.48$, which corresponds to the (quasi)balance between radioactive heating and convective cooling.
The increase in the heat flux $F$ during convective cooling is evident in Figure \ref{fig:OV_F}(c).

The initial spin period of Orcus is $3.1~{\rm h}$ in our simulation, and $P_{\rm p}$ increases with time (Figure \ref{fig:OV_F}(b)).
Since $k_{2}$ is sensitive to $T_{\rm c}$ (see Figure \ref{fig:k2}(a)), the time evolution of $P_{\rm p}$ varies significantly with $f_{\rm rock}$.
For $f_{\rm rock} = 0.48$, the Orcus--Vanth system reaches the synchronous state at $t \approx 2.1~{\rm Gyr}$, which is inconsistent with the current spin--orbit state \cite{2020EPSC...14..516K}.
The time evolution of $a$ is shown in Figure \ref{fig:OV_F}(d).
For $f_{\rm rock} = 0.24$, $a$ reaches approximately $8.6 \times 10^{3}~{\rm km}$ at $t = 4.5~{\rm Gyr}$, which is close to the observed value of the system \cite<$a_{\rm obs} = 9.0 \times 10^{3}~{\rm km}$;>[]{2019Icar..334...62G}.

\begin{figure}
\centering
\includegraphics[width = \columnwidth]{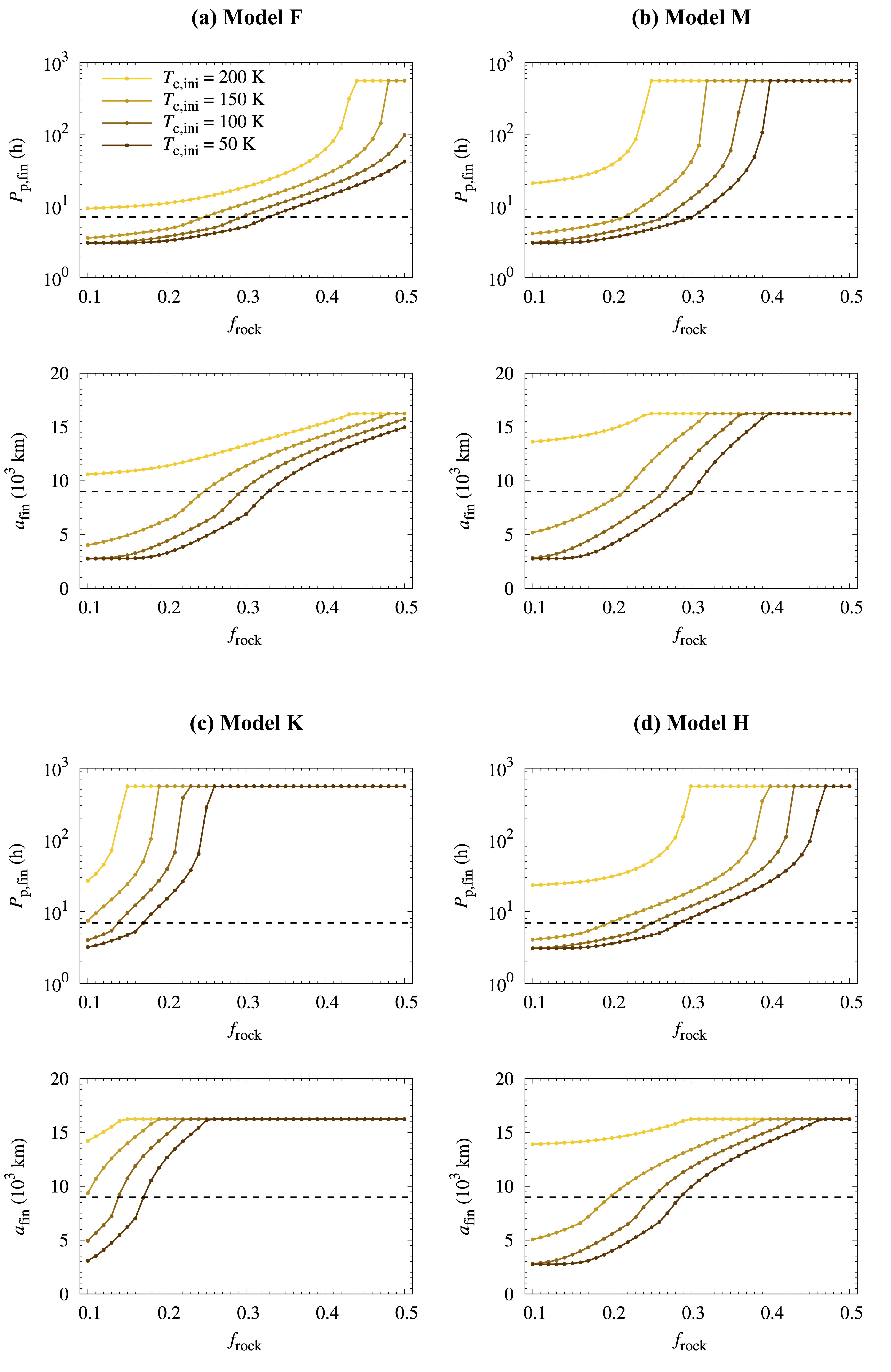}
\caption{
Spin period and the semimajor axis at $t = 4.5~{\rm Gyr}$ as functions of $f_{\rm rock}$ and $T_{\rm c, ini}$.
The upper panels show the spin period, $P_{\rm p, fin}$, while the lower panels show the semimajor axis, $a_{\rm fin}$.
(a) Model {\bf F}.
(b) Model {\bf M}.
(c) Model {\bf K}.
(d) Model {\bf H}.
}
\label{fig:OV_fin}
\end{figure}

Figure \ref{fig:OV_fin}(a) summarizes the spin period and the semimajor axis at $t = 4.5~{\rm Gyr}$ ($P_{\rm p, fin}$ and $a_{\rm fin}$) for Model {\bf F} as functions of $f_{\rm rock}$ and $T_{\rm c, ini}$.
The dashed lines represent the observed value for the current system: $P_{\rm p, obs} = 7~{\rm h}$ and $a_{\rm obs} = 9.0 \times 10^{3}~{\rm km}$.
We find that both $P_{\rm p, fin}$ and $a_{\rm fin}$ monotonically increase with $f_{\rm rock}$ and $T_{\rm c, ini}$.
Therefore, we can derive constraints on the ranges of $f_{\rm rock}$ and $T_{\rm c, ini}$ from Figure \ref{fig:OV_fin}(a): $f_{\rm rock}$ should be lower than $0.33$, and $T_{\rm c, ini}$ should be lower than $200~{\rm K}$.
If $T_{\rm c, ini}$ lies between $100~{\rm K}$ and $150~{\rm K}$, the observed spin--orbit state of the Orcus--Vanth system can be reproduced when $0.24 \le f_{\rm rock} \le 0.30$ for Model {\bf F}.

\begin{figure}
\centering
\includegraphics[width = \columnwidth]{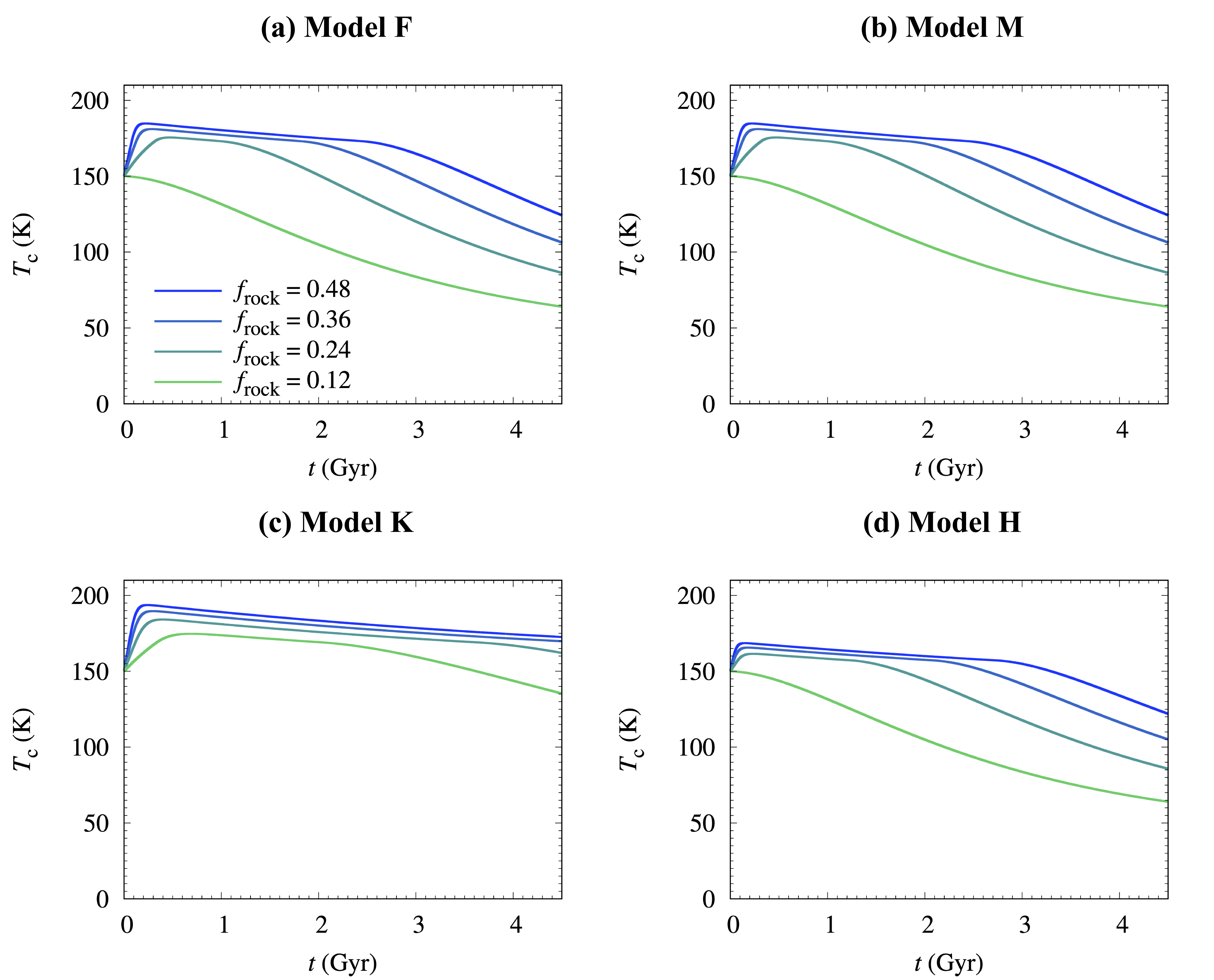}
\caption{
Thermal evolution of Orcus and its dependence on $f_{\rm rock}$ and the choice of models.
Here, we set $T_{\rm c, ini} = 150~{\rm K}$.
(a) Model {\bf F}. We note that this is identical to Figure \ref{fig:OV_F}(a).
(b) Model {\bf M}.
(c) Model {\bf K}.
(d) Model {\bf H}.
}
\label{fig:OV_T}
\end{figure}

We note that $P_{\rm p, fin}$ and $a_{\rm fin}$ depend on the choice of the model.
Figure \ref{fig:OV_fin}(b) shows the results for Model {\bf M}.
The thermal history of Orcus remains unchanged even if we choose a low $\mu$ value (see Figures \ref{fig:OV_T}(a) and \ref{fig:OV_T}(b)); however, ${\rm Im} {\left( k_{2} \right)}$ for Model {\bf M} is larger than that for Model {\bf F} when $T_{\rm c}$ is fixed.
Consequently, $P_{\rm p, fin}$ and $a_{\rm fin}$ for Model {\bf M} are larger than those for Model {\bf F}.
For $100~{\rm K} \le T_{\rm ini} \le 150~{\rm K}$, the observed spin--orbit state of the Orcus--Vanth system can be reproduced when $0.21 \le f_{\rm rock} \le 0.27$.

Figure \ref{fig:OV_fin}(c) shows the final spin--orbit state for Model {\bf K}.
As shown in Figure \ref{fig:k2}(c), ${\rm Im} {\left( k_{2} \right)}$ is nearly independent of $k_{\rm th}$.
In contrast, the thermal history of Orcus changes drastically when we set a low $k_{\rm th}$ (Figure \ref{fig:OV_T}(c)).
For $T_{\rm c, ini} = 150~{\rm K}$, the maximum $T_{\rm c}$ could exceed $170~{\rm K}$ even if $f_{\rm rock}$ is as small as $0.12$.
To reproduce the observed spin--orbit state, $f_{\rm rock} < 0.14$ is required for Model {\bf K} when $T_{\rm c, ini} \ge 100~{\rm K}$.

The dependence of the final spin--orbit state on $\eta_{\rm ref}$ is non-trivial, as discussed in \citeA{2021AJ....162..226A}.
When $T_{\rm c}$ is fixed, ${\rm Im} {\left( k_{2} \right)}$ increases with decreasing $\eta_{\rm ref}$ (Figure \ref{fig:k2}(d)).
However, $T_{\rm c}$ decrease with decreasing $\eta_{\rm ref}$ when convection controls the thermal evolution (see Figures \ref{fig:OV_T}(a) and \ref{fig:OV_T}(d)).
Our simulations find that $P_{\rm p, fin}$ and $a_{\rm fin}$ for Model {\bf H} are larger than those for Model {\bf F}, although this dependence is weaker compared to that for $k_{\rm th}$ (see Figures \ref{fig:OV_fin}(a) and \ref{fig:OV_fin}(d)).
For $100~{\rm K} \le T_{\rm ini} \le 150~{\rm K}$, the observed spin--orbit state of the Orcus--Vanth system can be reproduced when $0.19 \le f_{\rm rock} \le 0.26$.

\subsection{Salacia--Actaea}
\label{sec:SA}

\begin{figure}
\centering
\includegraphics[width = \columnwidth]{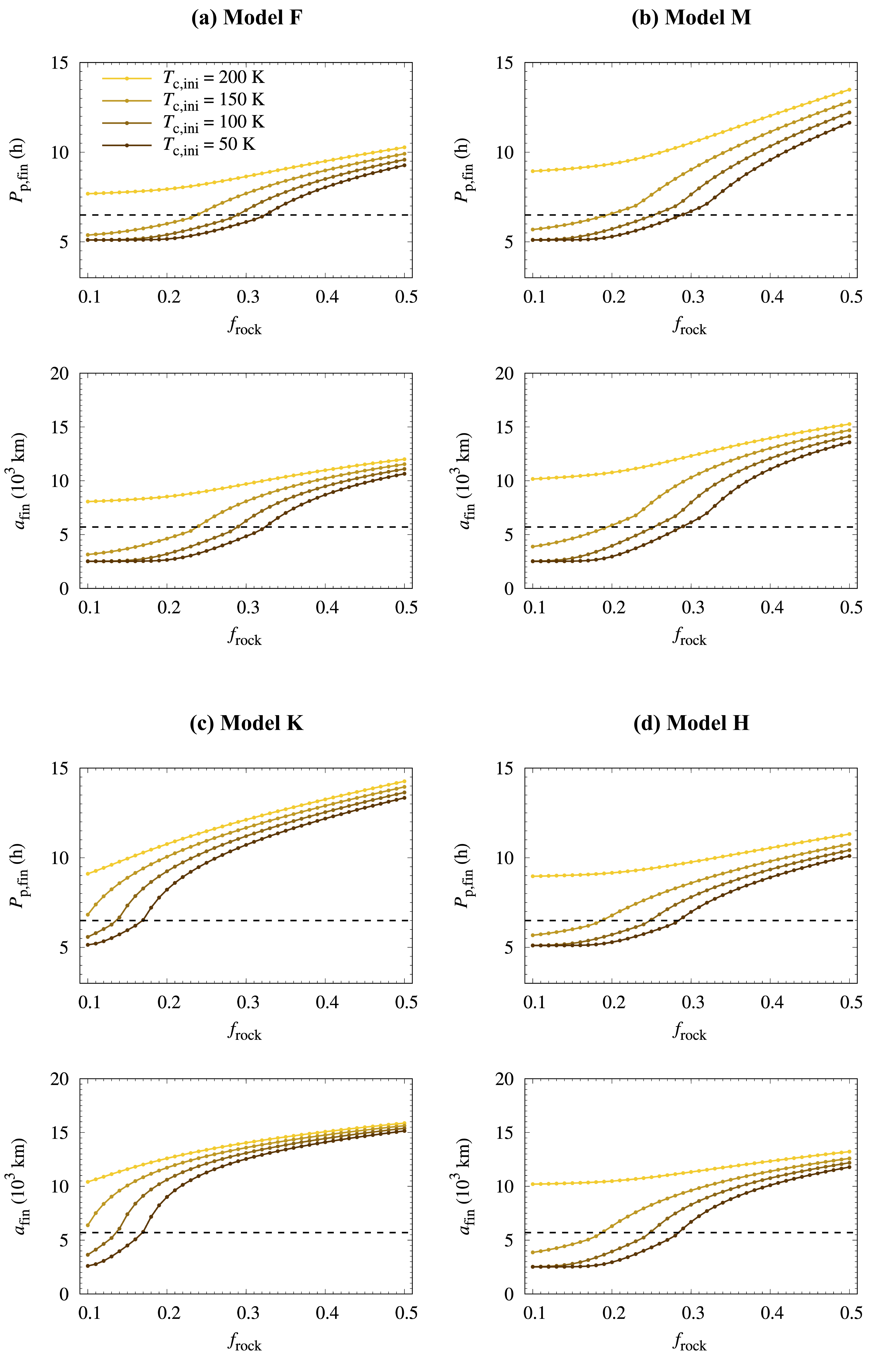}
\caption{
Same as Figure \ref{fig:OV_fin}, but for the case of Salacia--Actaea.
}
\label{fig:SA_fin}
\end{figure}

As in the case of Orcus--Vanth, we perform coupled thermal--orbital evolution calculations of the Salacia--Actaea system.
Figure \ref{fig:SA_fin} shows the final spin--orbit state of Salacia--Actaea for each model.
The upper panels show the final spin period of Salacia ($P_{\rm p, fin}$), while the lower panels show the final semimajor axis of the system ($a_{\rm fin}$).
The dependence on the model parameters is qualitatively the same as for Orcus--Vanth (Section \ref{sec:OV}).
The range of $f_{\rm rock}$ that can reproduce the observed spin--orbit state ( $P_{\rm p, obs} = 6.5~{\rm h}$ and $a_{\rm obs} = 5.7 \times 10^{3}~{\rm km}$) is $0.23 \le f_{\rm rock} \le 0.30$ for Model {\bf F} with $100~{\rm K} \le T_{\rm ini} \le 150~{\rm K}$ (Figure \ref{fig:SA_fin}(a)).
This range of $f_{\rm rock}$ is similar to that for Orcus--Vanth ($0.24 \le f_{\rm rock} \le 0.30$, see Figure \ref{fig:OV_fin}(a)).
Combining the results from the four models, we conclude that $f_{\rm rock}$ should be lower than $0.33$ and $T_{\rm ini}$ should be lower than $200~{\rm K}$.
This conclusion is consistent with that for Orcus--Vanth.

The estimated $f_{\rm rock}$ values for the Orcus--Vanth and Salacia--Actaea systems are approximately equal, suggesting that they share a similar chemical composition.
Furthermore, except for Eris, most large and mid-sized TNOs align with the size--density relationship shown in Figure \ref{fig:size-density}.
These observations may imply that TNOs formed within a common reservoir.

\section{Discussion}

\subsection{Volume and mass fractions of organic materials}
\label{sec:organic}

As shown in Section \ref{sec:results}, the observed spin--orbit states of the Orcus--Vanth and Salacia--Actaea systems suggest that $f_{\rm rock}$ is likely below $0.33$ for these two TNOs.
Additionally, based on the size--density relationship of TNOs with $R_{\rm p} > 100~{\rm km}$ (Figure \ref{fig:size-density}), the average material density is estimated to be $\rho_{\rm mat} \approx 1800~{\rm kg}~{\rm m}^{-3}$.
In this section, we evaluate the plausible ranges of the mass fractions of ice and organics ($f_{\rm ice}$ and $f_{\rm org}$) within TNOs.

\begin{figure}
\centering
\includegraphics[width = \columnwidth]{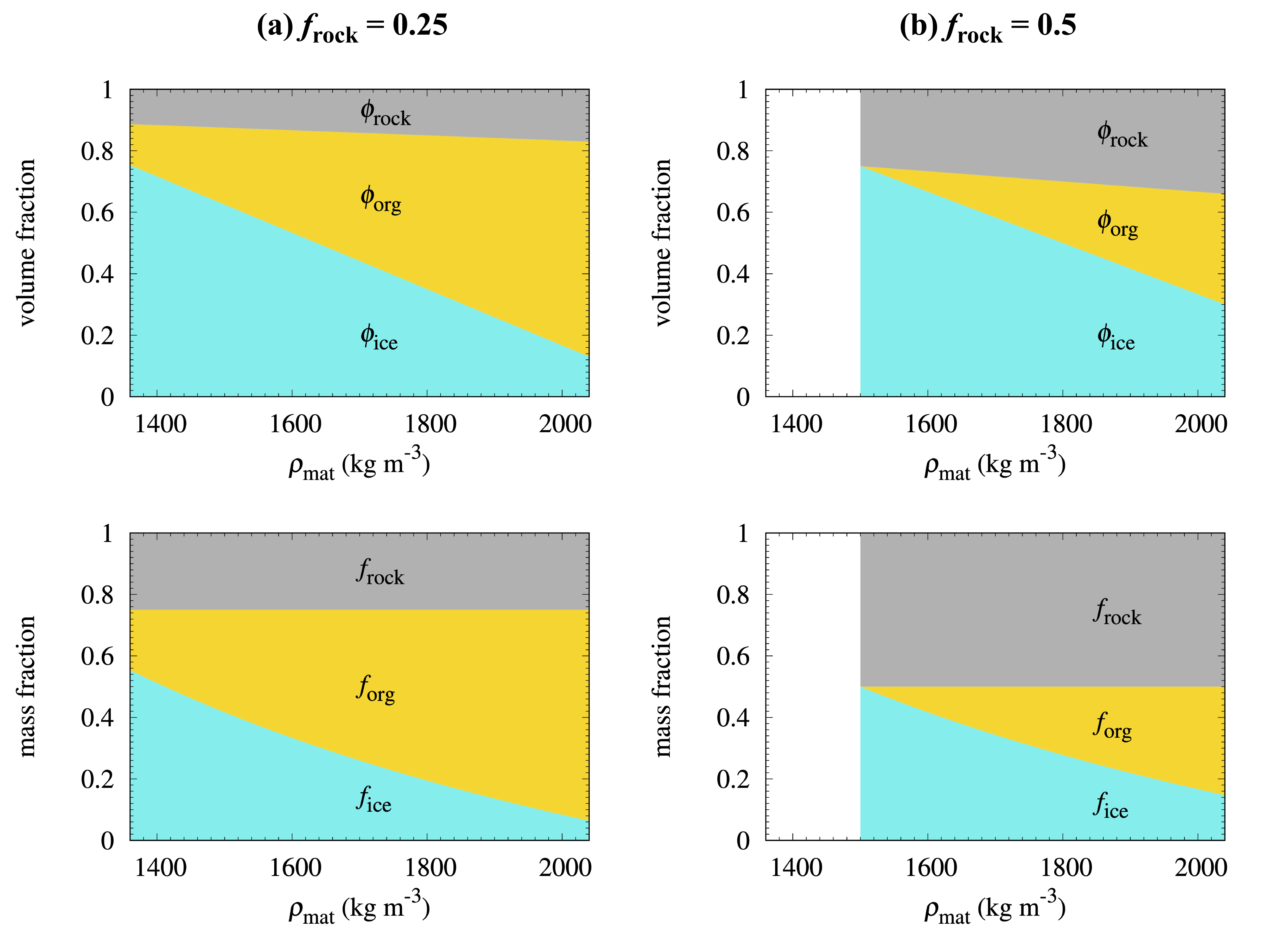}
\caption{
Volume and mass fractions of ice, organics, and rock ($\phi_{i}$ and $f_{i}$, where $i$ denotes ``ice'', ``org'', or ``rock'') as functions of $\rho_{\rm mat}$ and $f_{\rm rock}$.
(a) For $f_{\rm rock} = 0.25$.
(b) For $f_{\rm rock} = 0.5$.
}
\label{fig:phi_f}
\end{figure}

We define the volume fraction of the component $i$ (``ice'', ``org'', or ``rock'') as $\phi_{i}$, and its material density as $\rho_{i}$.
The volume fractions satisfy $\phi_{\rm ice} + \phi_{\rm org} + \phi_{\rm rock} = 1$, and $\rho_{\rm mat}$ is given by:
\begin{equation}
\rho_{\rm mat} = \phi_{\rm ice} \rho_{\rm ice} + \phi_{\rm org} \rho_{\rm org} + \phi_{\rm rock} \rho_{\rm rock}.
\label{eq:rho_mat}
\end{equation}
We assume $\rho_{\rm ice} = 1000~{\rm kg}~{\rm m}^{-3}$, $\rho_{\rm org} = 2000~{\rm kg}~{\rm m}^{-3}$, and $\rho_{\rm rock} = 3000~{\rm kg}~{\rm m}^{-3}$.
Our choice of $\rho_{\rm org} = 2000~{\rm kg}~{\rm m}^{-3}$ lies between the densities of coals ($\sim 1400~{\rm kg}~{\rm m}^{-3}$) and graphite ($\sim 2300~{\rm kg}~{\rm m}^{-3}$), which are analogues of macromolecular insoluble organic matter and amorphous carbons found in meteorites and interplanetary dust particles \cite<e.g.,>{2023E&PSL.61218172R}.
The mass fraction $f_{i}$ for each component is expressed as:
\begin{equation}
f_{i} = \frac{\rho_{i}}{\rho_{\rm mat}} \phi_{i}.
\label{eq:f_i}
\end{equation}
Using Equations (\ref{eq:rho_mat}) and (\ref{eq:f_i}), we calculate $\phi_{\rm org}$ and $f_{\rm org}$ as functions of $\rho_{\rm mat}$ and $f_{\rm rock}$, with results shown in Figure \ref{fig:phi_f}.

Figure \ref{fig:phi_f}(a) illustrates the volume and mass fractions of components for $f_{\rm rock} = 0.25$.
Based on spin--orbit evolution considerations, Orcus and Salacia likely share a rock mass fraction of $f_{\rm rock} \approx 0.25$.
Assuming an average material density of $\rho_{\rm mat} = {( 1800 \pm 150 )}~{\rm kg}~{\rm m}^{-3}$ (see Figure \ref{fig:size-density}), the corresponding range of organic mass fractions is $f_{\rm org}$ is $0.45 \le f_{\rm org} \le 0.64$.
For comparison, Figure \ref{fig:phi_f}(b) shows the volume and mass fractions of components for $f_{\rm rock} = 0.5$.
In this case, assuming the same range of $\rho_{\rm mat}$, the resulting organic mass fraction is $0.12 \le f_{\rm org} \le 0.31$.


The bulk densities of large TNOs suggest their organic-rich nature.
Recently, \citeA{2024ApJ...976...14T} demonstrated that their bulk densities can be explained by a comet-like, organic-rich composition, which is also consistent with the solar elemental abundances.
Organic materials are known to be abundant in comets.
For example, \citeA{2017MNRAS.469S.712B} reported an organic mass fraction of $f_{\rm org} \approx 0.45$ for dust particles ejected from comet 67P/Churyumov--Gerasimenko.
Similarly, \citeA{2022MNRAS.513.3734H} surveyed the ice--organic--silicate mass ratios of numerous comets and C-type asteroids, proposing that primordial comets are organic-rich bodies with mass fractions approximately ${\left( f_{\rm ice}, f_{\rm org}, f_{\rm rock} \right)} \approx {\left( 0.4, 0.3, 0.3 \right)}$.
The estimated $f_{\rm org}$ values from these studies roughly align with the results of our estimations.
However, it is important to note that the estimates of $f_{i}$ and $\phi_{i}$ are sensitive to the assumed material densities $\rho_{i}$, which remain uncertain \cite<see Section 2.3 of>[]{2020MNRAS.497.1166A}.

A large $f_{\rm org}$ value is also supported by the geochemical evolution of the largest TNOs with $R_{\rm p} > 500~{\rm km}$ \cite<e.g.,>[]{2019NatGe..12..407K, 2024Icar..41215999G}.
For instance, the coexistence of a subsurface ocean and a 1000-km-wide basin on Pluto suggests the presence of an insulating layer between the ocean and the ice shell.
Methane hydrates are considered the leading candidates for this insulator \cite{2019NatGe..12..407K}.
Additionally, recent James Webb Space Telescope observations of Eris and Makemake have revealed that surface methane ice on these dwarf planets is not primordial but may originate from the decomposition of organic materials \cite{2024Icar..41115923G, 2024Icar..41215999G}.
This finding implies that Pluto's methane hydrate layer could naturally form through thermally induced decomposition of organics.
Methane is the second most abundant molecule in the atmospheres of Pluto and Triton, and photochemical products such as ethylene and acetylene have also been clearly detected \cite<e.g.,>[]{2016Sci...351.8866G, 2019AREPS..47..119G}.
These observations suggest that the chemical and thermal states of their atmospheres and surfaces may provide critical insights into the thermal processes occurring in their interiors.

It should be noted that a large $f_{\rm org}$ value may influence the material parameters such as $\mu$ and $k_{\rm th}$.
Both the shear modulus and thermal conductivity of organic materials vary significantly depending on their composition and temperature.
Additionally, the viscosity of organics could become a key parameter if it is comparable to or even lower than that of ice.
However, there is currently limited information on the mechanical and thermal properties of organics in planetary bodies.
Laboratory experiments are crucial to quantitatively assess the effects of organics on the evolution of TNOs.

\subsection{Initial temperature of Orcus and Salacia}

Our secular spin--orbit evolution simulations indicate that $T_{\rm ini}$ for both Orcus and Salacia must have been lower than $200~{\rm K}$.
In this section, we examine the accretion history of these TNOs using analytic arguments constrained by the $T_{\rm ini}$ condition.

We consider two types of heat sources: the release of gravitational potential energy and the decay heat of radioactive isotopes.
For undifferentiated bodies with radius $R$ and density $\rho$, the gravitational potential energy, $U_{\rm grav}$, is given by $U_{\rm grav} = - {( 16 \pi^{2} / 15 )} \mathcal{G} \rho^{2} R^{5}$.
The maximum temperature increase due to the release of gravitational potential energy, ${( \Delta T )}_{\rm grav}$, is expressed as:
\begin{equation}
{( \Delta T )}_{\rm grav} = - \frac{U_{\rm grav}}{{( 4 \pi / 3 )} \rho R^{3} c} = 40~{\left( \frac{R}{400~{\rm km}} \right)}^{2}~{\left( \frac{\rho}{1500~{\rm kg}~{\rm m}^{-3}} \right)}~{\rm K},
\end{equation}
where $c$ is the specific heat capacity.
When the fraction of energy converted to heat is $\epsilon$, the temperature increase becomes $\epsilon {( \Delta T )}_{\rm grav}$.
During the growth of TNOs by the accretion of planetesimals, $\epsilon$ can be on the order of unity \cite<e.g.,>[]{2008ssbn.book..213M}, while for the accretion of tiny dust particles, $\epsilon$ is lower because a large fraction of the energy escapes as radiation \cite<e.g.,>[]{2008Icar..198..163B}.
Our analysis suggests that the release of gravitational potential energy would not cause a significant increase in $T_{\rm ini}$.
This result aligns with our conclusions from spin--orbit evolution simulations.

When planetesimals have random velocities, their impact velocity exceeds the escape velocity during accretion.
During the epoch of accretion in the primordial Kuiper belt, the random velocities among TNOs, $v_{\infty}$, are estimated to have typically been below $0.1~{\rm km}~{\rm s}^{-1}$, based on their growth timescales \cite<e.g.,>{2005Sci...307..546C}.
The maximum temperature increase due to the release of kinetic energy associated with random velocities, ${( \Delta T )}_{\rm kin}$, is given by ${( \Delta T )}_{\rm kin} = {( 1/2 )} {v_{\infty}}^{2} / c \approx 5 {( v_{\infty} / 0.1~{\rm km}~{\rm s}^{-1})}^{2}~{\rm K}$.
Thus, the contribution of kinetic energy to temperature increase would be negligible compared to the release of gravitational potential energy, although it could become significant if $v_{\infty}$ exceed $0.5~{\rm km}~{\rm s}^{-1}$.

Another heat source is the decay heat of radioactive isotopes.
If a TNO formed within the first few million years of the solar system, the decay of $^{26}{\rm Al}$ would dominate as the primary heat source \cite<e.g.,>[]{2017NatAs...1E..31S}.
The decay heat of $^{26}{\rm Al}$ per unit mass of rock at the time of the formation of calcium--aluminum-rich inclusions is $H = 1.9 \times 10^{-7}~{\rm W}~{\rm kg}^{-1}$, and the half-life time of $^{26}{\rm Al}$ is $\tau = 0.7~{\rm Myr}$ \cite{2021MNRAS.505.5654D}.
Assuming all decay heat is retained within the TNO, the temperature increase due to $^{26}{\rm Al}$ decay, ${( \Delta T )}_{\rm Al}$, is expressed as:
\begin{equation}
{( \Delta T )}_{\rm Al} = \frac{f_{\rm rock} H \tau}{c \ln{2}} \exp{\left( - \frac{t_{\rm acc}}{\tau / \ln{2}} \right)},
\end{equation}
where $t_{\rm acc}$ is the time of the accretion.
For $t_{\rm acc} = 2.5~{\rm Myr}$ and $f_{\rm rock} = 0.25$, we find ${( \Delta T )}_{\rm Al} = 127~{\rm K}$.
Given that $T_{\rm ini} < 200~{\rm K}$ is inferred from our spin--orbit evolution simulations, Orcus and Salacia likely formed at least $2.5~{\rm Myr}$ after calcium--aluminum-rich inclusions.

\subsection{Caveats in our current model}

We acknowledge certain caveats in this study.
To apply our simplified thermal evolution model, we have made several assumptions.
Specifically, the radial temperature profile is not calculated directly but is instead prescribed as a linear function of $r$ (Equation (\ref{eq:T_r})).
Since temperature-dependent rheology is a key parameter that governs spin--orbit evolution (Figure \ref{fig:k2}), simulations of spin--orbit evolution simulations incorporating precise temperature structure calculations would improve our understanding of the evolution of trans-Neptunian satellite systems.
However, fully three-dimensional simulations of thermal evolution are computationally intensive and not well-suited for parameter surveys.
For differentiated bodies, the radial temperature structure can be reasonably approximated using one-dimensional simulations based on (modified) mixing length theory \cite{2018JGRE..123...93K}, which could be coupled with tidal evolution calculations.
Although the modified mixing length theory for undifferentiated bodies is still under development \cite<e.g.,>[]{2022GeoJI.229..328V}, future studies coupling thermal and orbital evolution using one-dimensional thermal calculations would be essential for more quantitative discussions on the evolution of trans-Neptunian satellite systems.

We considered a spherical body for both thermal and orbital evolution calculations; however, the actual shapes of TNOs are non-spherical \cite<e.g.,>[]{2024AJ....167..144P, 2024PSJ.....5...69P}.
The $k_{2}$ value for non-spherical bodies is larger than that for spherical ones, but the difference is by a factor of only a few or less \cite<e.g.,>[]{2016MNRAS.463.1543Q}.
Therefore, we conclude that the uncertainty associated with non-sphericity would be smaller than that related to thermal history and temperature-dependent rheology.

We assumed that $R_{\rm p}$ and $\rho$ are constant over time.
In reality, however, both Orcus and Salacia likely experienced compaction through viscous relaxation of ice \cite<e.g.,>[]{2019Icar..326...10B}.
This densification would lead to an increase in $k_{\rm th}$ and $\mu$, potentially resulting in a decrease in $a_{\rm fin}$.
Since the timescale of densification is proportional to viscosity \cite{2019Icar..326...10B}, the final porosity of TNOs would significantly depend on the assumed values of $\eta_{\rm ref}$ and $k_{\rm th}$.
Thus, the size--density relationship of TNOs could provide a unique constraint on the thermal and mechanical properties of solids in the outer solar system.

We acknowledge that there is no direct evidence supporting our assumption of equal densities between the primaries and secondaries.
If both the Orcus--Vanth and Salacia--Actaea systems were formed via giant impacts between two undifferentiated bodies with nearly equal densities, we would expect the resulting primary and secondary to also have nearly equal densities.
In contrast, if the two colliding bodies were differentiated or had different densities before the impact, the densities of the primary and secondary formed by the giant impact could differ significantly.
We also note that $a_{\rm fin}$ is insensitive to the density of the satellite, $\rho_{\rm s}$, as $a_{\rm fin}$ is proportional to ${\rho_{\rm s}}^{2/13}$ \cite<e.g.,>[]{1999ssd..book.....M}.
Therefore, the impact of the density contrast on the spin--orbit evolution would be small.

The impact of the possible presence of amorphous ice on thermal evolution has been investigated by several authors \cite<e.g.,>[]{1999AdSpR..23.1299S, 2021MNRAS.505.5654D, 2024PASJ...76..130A}.
Since the thermal conductivity of amorphous ice is notably lower than that of crystalline ice, the presence of amorphous ice would reduce the $k_{\rm th}$ value, leading to an increase in $T_{\rm c}$, as discussed in Section \ref{sec:OV} (see Figures \ref{fig:OV_T}(a) and \ref{fig:OV_T}(c)).
Moreover, the crystallization of amorphous ice is an exothermic process \cite<e.g.,>[]{1968JChPh..48..503G}, which could cause an instantaneous increase in the local temperature $T {( r )}$ by several tens of kelvins.
The crystallization temperature of amorphous ice is $80$--$100~{\rm K}$ \cite<e.g.,>[]{2024PASJ...76..130A}, which is sufficiently higher than $T_{\rm s}$.
Thus, amorphous ice can persist in at least part of the conductive lid, where it could act as a thermal insulator.

\section{Conclusions}

In this study, we performed numerical simulations of the coupled thermal--orbital evolution of two trans-Neptunian satellite systems, Orcus--Vanth and Salacia--Actaea, using a viscoelastic rheological framework.
The viscoelastic response of the bodies, which depends on temperature, is a key parameter in the tidal evolution of satellite systems.
Since the primary heat sources of TNOs are radioactive nuclei, their internal temperature is governed by the initial abundances of these nuclei, which are proportional to the rock mass fraction within the objects.
By simulating coupled thermal--orbital evolution, we can therefore place constraints on the rock mass fraction within TNOs.
This study is the first to constrain the composition of TNOs based not only on their densities but also on their spin--orbit evolution.

We found that the current spin--orbit states of these two systems can be reproduced when the rock mass fraction is approximately 20--30\%, and the initial internal temperature of these TNOs is below $200~{\rm K}$ (see Figures \ref{fig:OV_fin} and \ref{fig:SA_fin}).
Additionally, we estimated the organic mass fraction within TNOs and found that it may be comparable to or even exceed the rock mass fraction (see Figure \ref{fig:phi_f}).
Interestingly, similar conclusions have been drawn in recent studies based on the bulk densities of TNOs \cite<e.g.,>[]{2024ApJ...976...14T}.
Our results underscore the potentially significant role of organic materials in the formation and evolution of planetary objects in the outer solar system.

\section*{Data Availability Statement}

The numerical code, {\bf LNTools}, is freely accessible at \citeA{2023zndo...7804175K}.

\acknowledgments

The authors thank anonymous reviewers for careful reviews and constructive comments.
This study was supported by JSPS KAKENHI Grant (JP24K17118).

\clearpage

%
\bibliography{agusample}
%


%
%
%
%
%

\end{document}